\documentclass[10pt]{iopart}
\expandafter\let\csname equation*\endcsname\relax
\expandafter\let\csname endequation*\endcsname\relax 

\usepackage{xstring}
\usepackage{amsmath}
\usepackage{graphicx}
\usepackage{dcolumn}
\usepackage{bm}
\usepackage{hyperref}
\usepackage[draft]{pgf}
\usepackage{wasysym}
\usepackage{amssymb}
\usepackage{siunitx}
\DeclareSIUnit\gauss{G}
\usepackage{tensor}
\usepackage{mathrsfs}  
\usepackage{dsfont}
\usepackage{mathtools}
\usepackage{xcolor}
\usepackage{booktabs,multirow}
\usepackage{xcolor}
\usepackage{cite}
\usepackage{epstopdf}

\usepackage{floatrow}
\newfloatcommand{capbtabbox}{table}[][\FBwidth]

\newcommand{\nextsetting}{\setting{*}}
\newcommand{\setting}[1]{\ensuremath{\mathbf{X}_{#1}}}

\newcommand{\E}{\mathrm{E}}
\newcommand{\Var}{\mathrm{Var}}

\renewcommand{\emph}[1]{\textcolor{red}{#1}}

\begin{document}
	
	\title[]{Applying machine learning optimization methods to the production of a quantum gas}
	
	\author{A J Barker$^1$, H Style$^1$, K Luksch$^1$, S Sunami$^1$, D Garrick$^1$, F Hill$^2$, C J Foot$^1$ and E Bentine$^1$}
	
	\address{$^1$ Clarendon Laboratory, University of Oxford, Parks Road, Oxford OX1 3PU, United Kingdom}
	\address{$^2$ DeepMind, 6 Pancras Square, London, N1C 4AG, United Kingdom.}
	\ead{adam.barker@physics.ox.ac.uk}
	\vspace{10pt}
	\begin{indented}
		\item[]\today
	\end{indented}
	
	\begin{abstract}
		
		We apply three machine learning strategies to optimize the atomic cooling processes utilized in the production of a Bose-Einstein condensate (BEC). For the first time, we optimize both laser cooling and evaporative cooling mechanisms simultaneously. We present the results of an evolutionary optimization method (Differential Evolution), a method based on non-parametric inference (Gaussian Process regression) and a gradient-based function approximator (Artificial Neural Network). Online optimization is performed using no prior knowledge of the apparatus, and the learner succeeds in creating a BEC from completely randomized initial parameters. Optimizing these cooling processes results in a factor of four increase in BEC atom number compared to our manually-optimized parameters. This automated approach can maintain close-to-optimal performance in long-term operation. Furthermore, we show that machine learning techniques can be used to identify the main sources of instability within the apparatus.
				
	\end{abstract}
		
	\vspace{2pc}
	

\section{Introduction}

Recent developments in artificial intelligence and machine learning have provided tools with which a computer can now outperform the analytic capability of a human, particularly when data sets are large or when a system relies on many free parameters~\cite{Patterson}. The application of machine learning methods has led to dramatic advances in many scientific fields and contexts, such as supply chain forecasting and healthcare~\cite{Min2010,Kermany2018}.
Machine learning is also well suited to the optimization of a complex experimental apparatus~\cite{Wigley2016,Tranter2018,Seif2018}.
As compared to a human, a major advantage of many machine learning methods is that the chosen learner has no preconceptions for how the parameters should affect the final result, and is therefore objectively guided purely by the actual data. As a result, a machine learner is able to find counter-intuitive solutions that a trained experimentalist may overlook~\cite{Tranter2018}.

In this paper, we apply three different machine learning algorithms to optimize an atomic physics experiment. 
Our apparatus is designed to produce a Bose-Einstein condensate (BEC), a quantum-mechanical state of matter which occurs when bosonic particles accumulate in their lowest energy (ground) quantum state~\cite{Einstein1925}. Bose-Einstein condensation in a dilute atomic vapour was first realized in 1995, resulting in the award of the Nobel Prize in 2001~\cite{Davis1995, CornellBEC1}. Since then, ultracold atomic vapour experiments have been used to investigate a wide range of physical phenomena, including quantum many-body physics~\cite{Bloch2012}, quantum-mechanical phase transitions~\cite{Hadzibabic2006,Greiner2002} and superfluid turbulence~\cite{Navon2016}.

To observe the BEC phase transition in dilute gas experiments, extremely low temperatures of tens of nanokelvin are typically required. The techniques used to reach these ultracold temperatures usually include a combination of optical cooling and forced evaporative cooling~\cite{Davis1995, Bradley1995, Anderson1995}. Implementing these cooling processes requires the precise sequencing of time-varying magnetic and optical fields using a control computer.
We parametrize these fields by defining their values at specific times, and refer to these definitions as the `settings' that describe a given sequence.
The parameter space that describes a typical experimental sequence is large and locating the optimal experimental settings using exhaustive, brute-force searches is unfeasible. 

Given the large parameter space, analytic models are often used to predict the optimal experimental settings. Well-established theory exists to explain several of the typical stages common to cold-atom apparatuses. For example, the cooling of atoms by the radiation forces exerted by laser light has been investigated for decades~\cite{Cohen-Tannoudji1998} and forced evaporative cooling in optical or magnetic traps is routinely used~\cite{Ketterle1996}. The theories describing these stages of cooling contain approximations. Furthermore, the apparatus can suffer from unknown imperfections or external perturbations. These limitations are usually mitigated by employing further manual optimization of the experimental settings after using the theoretical optimum predictions as a starting point. 

Recently, machine learning techniques have been applied in the field of ultracold quantum matter to optimize individual laser cooling~\cite{Tranter2018,Toscano2019} and evaporative cooling~\cite{Geisel2013,Wigley2016} stages, achieving significant improvements in the performance of these apparatuses. The optimizations in each case were performed on a subset of the atomic cooling processes~\cite{Lausch2016,Rohringer2008}, and did not consider the changeover between each process, which increases the likelihood that the entire cooling sequence will become trapped in a local optimum.

In this paper, we present the results of a simultaneous optimization of all atomic cooling stages involved in our experimental sequence. Additionally, we compare the efficacy and rate of convergence of three common algorithms when applied to our optimization problem. The exact nature of what constitutes an optimized quantum gas experiment depends on the user's requirements. For example: quicker experiments with a higher repetition rate produce a greater amount of data in a given time; lower temperatures of the atomic cloud can improve the precision of spectroscopic measurements~\cite{Harte2018}; a larger atom number or higher peak density can improve the signal-to-noise ratio when imaging the BEC. Here, our chosen metric for optimization consists of maximising the atom number in a BEC, unless stated otherwise. 

We define our methods of optimization in section~\ref{Sec:Opt_Methods} and implement these using an open-source software package (M-LOOP)~\cite{MLOOP}, which has previously been used to optimize evaporative cooling elsewhere~\cite{Wigley2016}. The improvements in experimental performance that result from the optimization of several cooling stages, both individually and collectively, are then presented. We utilize one particular optimization method to identify those experimental settings which most strongly affect the result~\cite{Wigley2016}; this also highlights likely sources of instability within the experiment. Finally, we modify our optimization metric to minimize the sequence time required to produce a BEC, which is desired when performing tasks such as optical alignment, or to collect more data when atom number is not a priority.

\section{The experimental apparatus}

\begin{figure}[h]
	\centering
		\includegraphics[width = 0.9\linewidth]{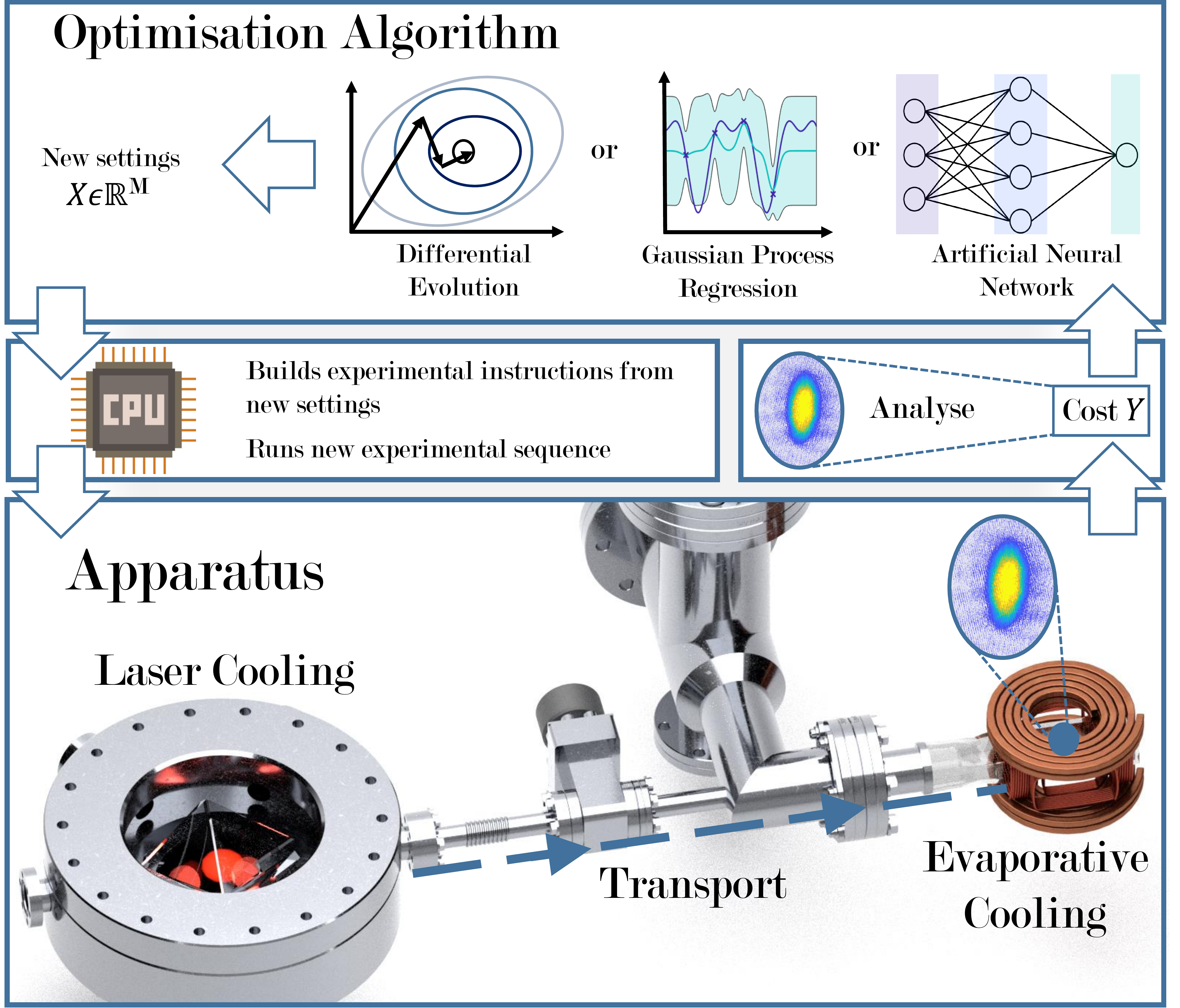}
	\caption{The experimental apparatus and the optimization loop. The atomic gas is initially trapped and laser cooled by a combination of laser light and magnetic fields. The trapped cloud is then transported to an ultra-high vacuum region where evaporative cooling is performed. An image of the resulting cloud is taken using absorption imaging~\cite{Foot2005} and is analysed to evaluate the cost, which is calculated from the atom number in the cloud. The cost is passed to the chosen optimization algorithm which produces a new set of experimental settings. These settings are processed by a control computer into a new set of instructions and used in the next sequence.}
	\label{fig:Full_Loop}
\end{figure}

We now describe our experimental apparatus and the several stages of trapping and cooling of an atomic vapour which lead to the production of a BEC~\cite{Harte2018, Gildemeister2010}. An outline of the apparatus and optimization scheme is illustrated in figure~\ref{fig:Full_Loop}. 

\subsection{Producing a Bose-Einstein condensate}
\label{Sec:evap}
First, atoms are laser cooled in a magneto-optical trap (MOT)~\cite{Raab1987}, which collects a fraction of atoms from a room temperature vapour and cools them to around the Doppler limit (\SI{146}{\micro \K} for ${}^{87}\mathrm{Rb}$) \cite{Steck2001}. After fully loading the MOT, the trapped atoms are subjected to a sudden compression and further cooling during a `compressed' MOT (cMOT) stage, which acts to further reduce the temperature by roughly an order of magnitude~\cite{Sherlock2011,Sheard2011}. The efficiency of the cooling and compression is dependent on many factors which include the detuning of the laser light from the atomic resonance and the strength of the applied magnetic field. The cold cloud is loaded into a magnetic quadrupole trap and transported to an ultra-high vacuum region by physically translating the field-producing coils. Subsequently, evaporative cooling is performed to further reduce the cloud temperature.

Evaporative cooling can be understood from the following arguments.
Atoms in a gas at a finite temperature occupy a distribution of energies, as described by the Maxwell-Boltzmann distribution~\cite{Blundell2010}. Evaporative cooling is performed by selectively ejecting the highest-energy atoms, which reduces the average energy of the remaining atoms. The trapped atoms then rethermalize through collisions, which re-establishes a Maxwell-Boltzmann distribution characterized by a lower temperature~\cite{PethickSmith}. In our case, evaporation is performed by the application of a weak radiofrequency (RF) field, colloquially referred to as a `knife'\footnote{The applied RF field effectively cuts away the high energy tail of the Maxwell-Boltzmann distribution, hence it is termed a `knife'.}, which removes atoms with energy above a threshold determined by the frequency of the applied knife.

Evaporation is first performed in a magnetic quadrupole trap and later in a time-averaged orbiting potential (TOP) trap~\cite{CornellBEC1, Bentine}. The quadrupole trap is implemented using a pair of coaxial current-carrying coils to produce a magnetic quadrupole field that confines the atoms. After the RF knife is applied to the trapped atoms, the frequency is slowly reduced; as the evaporation stage progresses, this reduces the threshold energy at which atoms are removed and thus reduces the cloud temperature. 

After a first stage of evaporative cooling, atoms are loaded into the TOP trap. This typically occurs once a temperature of \SI{1}{\micro \K} is reached.
The magnetic field of the TOP trap combines a static quadrupole field with a rotating bias field that lies in the horizontal plane, and is of the form $B = B_x \mathrm{cos} ( \omega t ) \hat{\textbf{e}}_x + B_y \mathrm{sin}(\omega t) \hat{\textbf{e}}_y$, where $\omega = 2\pi \times \SI{7}{\kilo \hertz}$ is the field rotation frequency,  $\hat{\textbf{e}}_{x,y}$ are the Cartesian axes in the horizontal plane and $B_x$ and $B_y$ are the amplitudes of the quadratures of the field. $B_x$ and $B_y$ can be individually controlled to produce an elliptically polarized field, with the ellipticity expressed by $\epsilon = B_y/B_x - 1$. Further evaporative cooling proceeds in the TOP trap using the RF knife as before. Overall, the evaporative cooling processes in both traps are described by a number of settings which vary in time and include: the quadrupole coil current $I_{\mathrm{Q}}$, the RF knife frequency and, in the case of the TOP trap, the amplitude and ellipticity of the TOP field. 

The experimental settings are processed by the control computer in order to direct the apparatus during the sequence. By adjusting these settings between successive sequences, we are able to optimize the production of a BEC.

\subsection{Observing a Bose-Einstein condensate}

\label{Sec:readout}
After all stages of cooling have been completed, the atomic cloud is released from the trap. The cloud undergoes a period of free fall, during which it expands ballistically, before an image is taken~\cite{Foot2005}. This `time-of-flight' (TOF) expansion allows us to observe the momentum distribution of the cloud. The expansion dynamics of a gas in the quantum regime are distinct from those of a thermal gas~\cite{PethickSmith}. This difference produces a bimodal spatial distribution of atoms after TOF: the BEC component is responsible for a dense `core' of atoms which lies within a broader `pedestal' of thermal atoms. This bimodal distribution is evidence that a BEC has been produced. The absorption image is analyzed to determine properties of the cloud, such as the atom number, which are used in the calculation of the cost.

\subsection{Machine learning methods}
\label{sec:ML_methods}
The goal of optimization is to identify the global optimum within a parameter space. In our experiment, the parameter space is that spanned by $M$ experimental settings (currents, voltages, timings etc.). A point in this parameter space is given by a vector of experimental settings $\mathbf{X} \in \mathbb{R}^M$ \footnote{Although the settings are continuously varying (up to floating-point precision), bounds on each setting are imposed, owing to physical limitations or for safety reasons, hence the set is not strictly $\mathbb{R}^M$.}. Each point in space has an associated cost $Y = f(\mathbf{X}) \in \mathbb{R}$, generated by a cost function $f(\mathbf{X})$~\cite{Glover2003}. The cost function quantifies the desirability of a measured outcome, and is used to steer the optimization. 

\begin{figure}
	\centering
	\includegraphics[width = 0.8\linewidth]{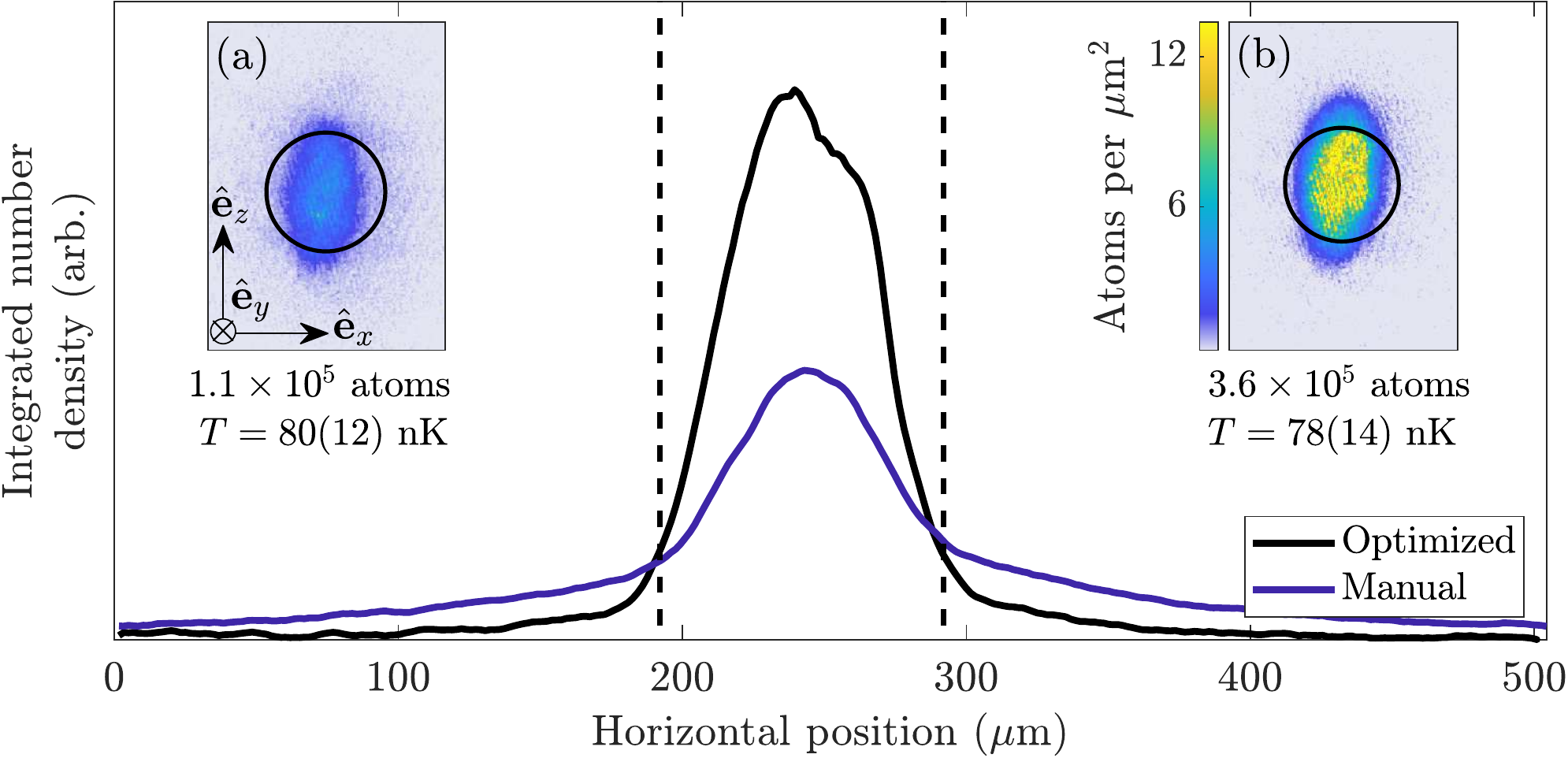}
	\caption{Illustration of the results of the optimization via machine learning versus manual optimization. We compare the absorption images, showing the atomic density integrated along the imaging direction $\hat{\textbf{e}}_{y}$, of a manually-optimized BEC (a) and a BEC where the evaporative cooling stages of both the quadrupole and TOP traps have been optimized using the GP method (b). Both inset images show the region (black line) in which the counted atom number contributes to the cost. The total atom numbers and temperatures of the clouds are shown below the images. The plot shows a direct comparison of the images integrated along the direction of gravity, $\hat{\textbf{e}}_{z}$. The horizontal extent of the region of interest, within which atoms are counted in the cost function, is indicated by the dashed lines.}
	\label{fig:Absorption}
\end{figure}

There are a number of candidate cost functions that could be used for a cold atom experiment. Figure~\ref{fig:Absorption} illustrates an example absorption image, from which a wealth of data may be obtained. The bimodal density distribution after TOF expansion clearly indicates the presence of a BEC, as explained in section~\ref{Sec:readout}. Consequently, we define our cost function to be proportional to $-\mathrm{log}(\tilde{N})$, where $\tilde{N}$ is the number of atoms within a small region of interest which is chosen to be comparable to the approximate extent of a typical BEC after TOF expansion~\cite{PethickSmith}. 
Atoms above a threshold momentum are not contained within this region after TOF expansion, and therefore do not contribute to $\tilde{N}$. Further detail justifying our choice of cost function is presented in~\ref{sec:app}.
We choose the cost function to be the logarithm of $\tilde{N}$, as the value of $\tilde{N}$ can span several orders of magnitude during the optimization; bad settings may result in no atoms detected above the noise floor, whereas BECs typically contain approximately $10^5$ atoms. 

Although analytic functions exist which describe the bimodal distribution, they contain many free parameters. This can make fitting unreliable, especially for the low atom numbers present in the early stages of optimization. Parameters extracted from such fits can throw the learner off course; our simple cost function is robust against these issues. Physically, our cost function can be interpreted as measuring the population of atoms with momentum close to zero, which increases as the optimization progresses towards producing a BEC.
Previous work employed a cost function derived from the fitted width of the cloud, with two repeats per experimental settings generation~\cite{Wigley2016}. 

The optimization feedback loop, outlined in figure~\ref{fig:Full_Loop}, can be summarized as follows: the machine learner is configured with an initial $M$-dimensional vector \setting{0} of experimental settings. We also configure the allowed ranges that each setting can take, to ensure that generated sequences will not damage the apparatus. \setting{0} is read by an experimental control computer, which defines relevant analog and digital outputs at time steps accordingly. The sequence is run and the resulting image is analyzed to produce a cost $Y_0$. This pair of settings and cost is then used by the chosen optimization algorithm to determine the next settings \nextsetting{} to be tested. Each new settings/cost pair updates the learner's knowledge of how the cost depends on each setting~\cite{Wagner}.
Our problem describes a settings/cost landscape with no initial data and is an example of online optimization. We terminate the optimization after a fixed number of sequences or when no further improvement to the cost has been achieved after 35 sequences. 

To implement the optimization routines, we utilize an open-source machine learning toolkit: Machine Learning Online Optimization Package (M-LOOP), which is based on the Python scikit-learn library~\cite{Wigley2016,MLOOP}. This toolkit contains several optimization routines, such as Gaussian Process regression and Differential Evolution, which are described in section~\ref{Sec:Opt_Methods}.

\section{Optimization methods}
\label{Sec:Opt_Methods}

We compare the efficacy of three algorithms to optimize our experiment: an evolutionary optimization method (Differential Evolution)~\cite{Storn1997}, a regression method based on non-parametric inference (Gaussian Process regression)~\cite{SEEGER2004} and a gradient-based (parametric) function approximator (Artificial Neural Network)~\cite{Patterson}. 

The optimization methods are tested in the context of non-convex optimization: the cost function described earlier is in general non-convex and thus it is possible that any method may not converge to the global optimum. The likelihood of finding the global optimum can be increased by performing many optimization procedures with varying initial conditions. 

We note that the optimization methods are robust to random variations in cost for a given setting. This is appropriate for an experimental apparatus in which random fluctuations are present, either due to variation in the performance of laboratory equipment or because the results depend intrinsically on random processes (e.g. shot noise fluctuations in the atom number). This does not fundamentally prevent the algorithms from finding a good solution, but uncertainty in the cost increases the number of experimental sequences required for the solution to converge.

\subsection{Differential Evolution}

Evolutionary algorithms involve several key stages, which are inspired by biological evolution~\cite{Vikhar2016}. First, an initial population is generated randomly. New individuals are then produced by mixing features of pre-existing individuals (crossover) and by adding random variation (mutation). Finally, selection is performed by assessing the fitness of new individuals and by replacing the population with the lowest fitness. 

In the present work, we use the Differential Evolution (DE) algorithm. In this context, the individuals are settings vectors \setting{i} and the fitness is the associated cost $Y_i$ of each vector. The initial population is a randomly generated set of $n$ vectors $\{\setting{1},...,\setting{n}\}$ and their experimentally measured costs $\{Y_1,...,Y_n\}$. Mutation produces a new vector $\mathbf{V} = \setting{k} + ( \setting{i} - \setting{j} )$, where \setting{i}, \setting{j} and \setting{k} are randomly selected vectors~\cite{Storn1997}. Crossover is achieved by selecting elements randomly from either \setting{i} or $\mathbf{V}$ to create a new candidate vector \setting{*}. A sequence is then performed using vector \setting{*} and the value of the cost function $Y_{*}$ is measured, producing an additional settings/cost pair $\{\setting{*}, Y_{*}\}$.
Selection is performed by determining whether $Y_* < Y_i$; if so, \setting{*} replaces \setting{i} and the process repeats. 

The DE method has low computational complexity and requires a small number of vectors from which to begin. However, the simplicity of the method results in slow convergence towards a solution. Nevertheless, we utilize this method to build a set of settings/cost pairs which serves as a starting point to initially train the other optimization methods.

\subsection{Gaussian Process regression}
\label{Sec:GP}

Bayesian inference provides us with tools to update a prior hypothesis of a probability distribution based on new data, namely Bayes' rule~\cite{SEEGER2004}. In general, a Gaussian Process (GP) is a probability distribution of functions which describe a given dataset. GP regression utilizes Bayes' rule to update this probability distribution given new data~\cite{Rasmussen2006}. Prior knowledge about a point in parameter space can be invoked in terms of a kernel function; a kernel is a measure of similarity between two inputs separated by a distance in parameter space. A popular choice is the squared-exponential, or Gaussian, distribution kernel $K(\setting{i}, \setting{j})$:

\begin{equation}
K(\setting{i}, \setting{j}) = \mathrm{exp}\left\{ -\frac{1}{2}\sum^{M}_{k=1} \eta_k(\setting{i}[k] - \setting{j}[k])^2 \right\} ,
\end{equation}

where $\setting{i}[k]$ represents the (dimensionless) $k$-th element in the vector \setting{i}, the dimensionless parameters $1/\eta_k$ are the characteristic length-scales for each parameter and the summation runs over all settings $k$. The $\eta_k$ are generated when performing GP regression, and provide a measure of how strongly the kernel depends on changes to each of the parameters. 

In our context, the function that we fit using GP regression is the mapping between the experimental settings and the experimentally measured cost. Given an existing set of settings/cost pairs $\{\setting{i}, Y_i\}$, we can estimate the cost (and uncertainty) of any settings \setting{*} according to the GP fit. We can therefore search for new experimental settings with the lowest predicted cost and iterate within our optimization loop. To facilitate a comparison of $\eta_k$ across all settings, we normalize each $\setting{}[k]$ with respect to the minimum and maximum allowed values for the $k$-th setting. Before applying the GP method, a training set of $2M$ settings/cost pairs is constructed using the DE method.

\subsection{Artificial Neural Network}

Artificial Neural Networks (ANN) are an example of a function approximator and take the form of an interconnected network of nodes~\cite{Patterson, Schmidhuber2015}. The ANN produces a `black-box' mapping between an input and an output. In our context, the inputs are settings vectors $\setting{}$ and the outputs are the associated costs $Y$. The mapping is determined by the structure and weights of connections in the network. This connection structure is intrinsically linear. To incorporate the non-linearity of the cost function, we include the Gaussian Error Linear Unit (GELU) activation function for each node~\cite{Hendrycks2016}. This continuous function is a popular choice for data which is subject to normally-distributed stochastic variation, which suits our experimental context~\cite{Kalantre2019}. In addition, the structure and scale of the ANN must be appropriate for the complexity and size of the vector inputs. We choose a network comprised of 3 hidden layers of 8 fully-connected neurons, inspired by~\cite{Sheela2013}, which is sufficient for the number of settings that we optimize in this context (a maximum of 35). An initial training set of $2M$ settings/cost pairs is produced using the DE method.

We utilize the Adam optimization method~\cite{Kingma2014} to update the ANN given new training data. This method is widely used for gradient-based optimization of cost functions with stochastic noise. The method is straightforward to implement and is computationally efficient; the method is also appropriate for problems with very noisy or sparse gradients. In comparison to other classical gradient descent methods, the Adam method utilizes higher-order moments of the gradients of each parameter~\cite{Ruder2016}, which often leads to a comparatively faster learning rate~\cite{Kingma2014}. We use the trained ANN to search for optimal predicted settings \setting{*}. A sequence is then run using \setting{*} and the cost $Y_{*}$ is measured. This settings/cost pair is then used to refine the ANN for future sequences.

\section{Results}

We applied the algorithms presented above to optimize the atomic cooling processes utilized in the production of BECs. We begin by presenting the optimization of evaporative cooling in the quadrupole and TOP traps. This optimization also identified the settings that most strongly affected the cost function. Similarly, we optimized the cMOT laser cooling stage. We combined the sensitive settings in both the laser cooling and evaporative cooling stages to perform a full optimization of all cooling processes involved in the production of a BEC. Finally, we altered the cost function to favour faster sequences, finding settings which produced a BEC of a threshold atom number within the shortest sequence duration. 

\subsection{Optimizing evaporative cooling}
\label{sec:evapCooling}
Prior to machine learning optimization, our manually-optimized settings produce a BEC of $1.1 \times 10^5$ atoms. This produces a cost of $8.9$ when using a circular region of interest of radius 50 $\mu$m located about the cloud centre after 23 \si{\milli \second} of free fall. These settings can be used as a starting point for machine learning optimization, leading to rapid convergence towards the optimum settings and providing a useful way to quickly retune the experiment. 

\begin{figure}[t]
	\centering
	\includegraphics[width = \linewidth]{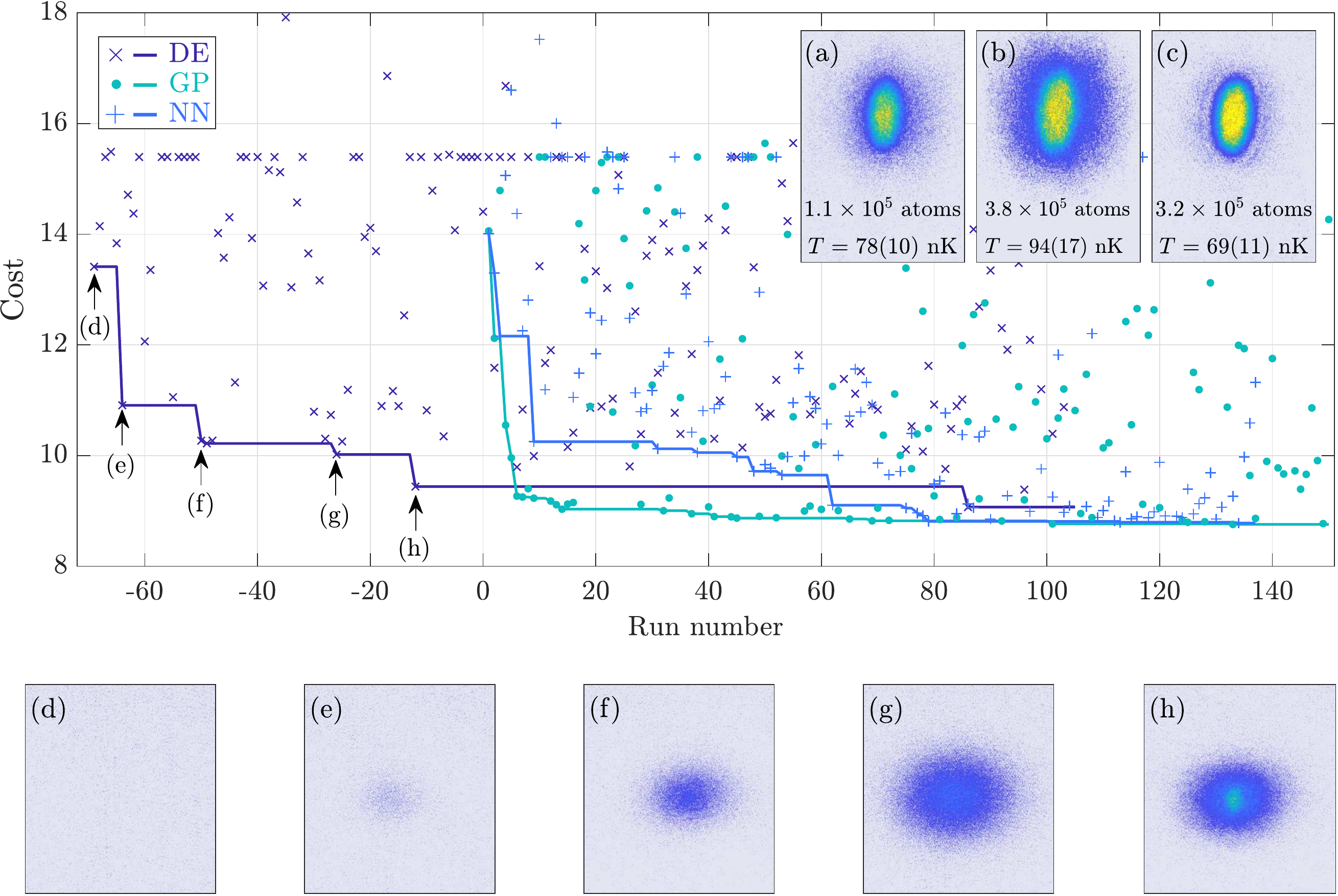}
	\caption{Optimizing the quadrupole and TOP evaporative cooling stages, beginning from random initial settings. Data points for the measured cost are illustrated as \textcolor[rgb]{0.2422,0.1504,0.6603}{\textbf{$\times$}} (DE), \textcolor[rgb]{0.0433,0.7411,0.7394}{$\bullet$} (GP) and \textcolor[rgb]{0.2206,0.4603,0.9973}{$+$} (ANN). In addition, the moving minimum for each of the three methods is indicated by the solid lines. Inset (a,b,c): absorption images of BECs produced using the best settings found for DE, GP and ANN-based optimization, respectively, including the atom number and temperature for each cloud. Sequences used in building the training set using the DE method are labelled with negative numbers. Absorption images below correspond to points labelled (d), (e), (f), (g) and (h). These images illustrate the progression of the optimization from initial settings, for which no trapped atoms were detected, towards finding settings which produce a BEC. The images use the same colour scheme as used in figure~\ref{fig:Absorption}.}
	\label{fig:Random_Opt}
\end{figure}

In order to properly compare the different learners, we instead begin each optimization using completely randomized settings. These initial settings produce no visible atom cloud. Fig.~\ref{fig:Random_Opt} shows the cost as a function of experimental run number. The optimization is continued until no further improvement is found within 35 cycles or until a maximum of 180 sequences, which limits the optimization process to a maximum duration of approximately 3 hours.
We perform one optimization routine for each method.

The GP method converged to a BEC of $3.8 \times 10^5$ atoms after 47 sequences, whereas DE did not converge within the time limit. The ANN method produced a BEC of $3.2 \times 10^5$ atoms after 117 sequences with a rate of convergence which is between those of the GP and DE methods \footnote{For both the GP and ANN methods, the quoted number of sequences does not include the training set of $2M=70$ sequences which was produced using DE.}. For both the GP and ANN methods, the optimization procedure resulted in a factor of 3 increase in BEC atom number as compared to the manually-optimized settings. The settings that produced the best cost are shown in table~\ref{fig:EvapArray} with the original settings used prior to optimization shown in parentheses. Figure~\ref{fig:Length_Scales} (a) \& (b) illustrate the progression of the experimental settings during the optimization, namely the quadrupole current $I_Q$ and RF knife frequency, respectively. The settings are plotted against the duration of each substage.

\begin{table}[htbp]
	\caption{\label{fig:EvapArray} The best settings found for evaporative cooling stages in the quadrupole (Quad) and TOP traps; these were found using the GP method but are very similar to those found using the ANN method. The values shown define points which are linearly interpolated to produce the evaporation instructions. Numbers in parentheses represent the manually-optimized settings used prior to the optimization. Values shown without brackets were not included in the optimization.}
	\centering
	\begin{tabular}{p{0.01\textwidth}p{0.1\textwidth}p{0.03\textwidth}p{0.07\textwidth}p{0.025\textwidth}p{0.075\textwidth}p{0.02\textwidth}p{0.08\textwidth}p{0.025\textwidth}p{0.063\textwidth}p{0.058\textwidth}p{0.05\textwidth}}  
		\toprule
		& Substage & \multicolumn{2}{p{1.5cm}}{Duration (s)} & \multicolumn{2}{p{1.5cm}}{$I_Q$ (A)} & \multicolumn{2}{p{1.5cm}}{RF knife (MHz)} & \multicolumn{2}{p{1.6cm}}{$B_x$ (G)} & \multicolumn{2}{p{1.5cm}}{Ellipticity, $\epsilon$} \\ 
		\midrule
		\parbox[t]{2mm}{\multirow{2}{*}{\rotatebox[origin=c]{90}{Quad}}} & 0 & 0 & & 323 & (315) & 120 & & & & & \\
		& 1 & 18 & & 323 & (315) & 15 & (18) & & & & \\
		\midrule
		\parbox[t]{2mm}{\multirow{7}{*}{\rotatebox[origin=c]{90}{TOP}}}
		& 2 & 0 & & 83 & (60) & 26 & (32) & 2.6 & (3.6) & 0 & (0)\\
		& 3 & 0.08 & (0.08) & 142 & (131) & 29 & (26) & 19 & (18) & 0 & (0)\\
		& 4 & 8.1 & (7.0) & 237 & (226) & 9.1 & (9) & 6.6 & (7.8) & 0.06 & (0)\\ 
		& 5 & 1.1 & (0.8) & 213 & (226) & 10 & (8.5) & 6.3 & (7.8) & -0.15 & (0)\\
		& 6 & 1.8 & (1.8) & 249 & (226) & 14 & (7.8) & 9.9 & (7.8) & 0.04 & (0)\\
		& 7 & 6.3 & (3.3)& 222 & (226) & 9.5 & (6.7) & 9.0 & (7.8) & 0.09 & (0)\\
		& 8 & 5.5 & (1.5) & 200 & (226) & 6.9 & (6.5) & 7.8 & (7.8) & 0.11 & (0)\\ \bottomrule	
	\end{tabular}
\end{table}

The cloud density profile after TOF expansion becomes bimodal as the cost function drops below approximately 9.2, indicating the presence of a BEC component. This threshold is achieved after 156 sequences (DE), 14 sequences (GP) and 75 sequences (ANN). Overall, the relative convergence rates of the methods differ significantly, as expected; the slower convergence of the ANN, as compared to GP, is representative of the large amount of data required to train a fully-connected network. DE proceeds the slowest of the three, which is expected given its simplistic approach to generating subsequent settings. 
As there is an element of randomness as to how the DE method chooses points to evaluate, it is possible that one model may have chanced upon good settings early in the optimization procedure which then strongly guided its subsequent choices.
However, the data sets used to train the GP and ANN methods were comparable and contained settings with minimum costs of 9.4 and 9.6, respectively, giving a fair comparison between the learners.

After only a few hours, the result of the optimization procedure is a greater than threefold improvement in the BEC atom number, as compared to a BEC produced with settings which have been manually optimized over many years. This improvement is illustrated in figure~\ref{fig:Absorption}, (a) \& (b), which show BECs produced using the evaporation settings before and after optimization, respectively. Figure~\ref{fig:Absorption} also demonstrates the merit of defining a small region of interest, which discounts atoms from the broader, thermal fraction of the atomic distribution. 

\subsection{Sensitivities}
\label{Sec:Sensitivites}

\begin{figure}
	\centering
	\includegraphics[width = 0.95\linewidth]{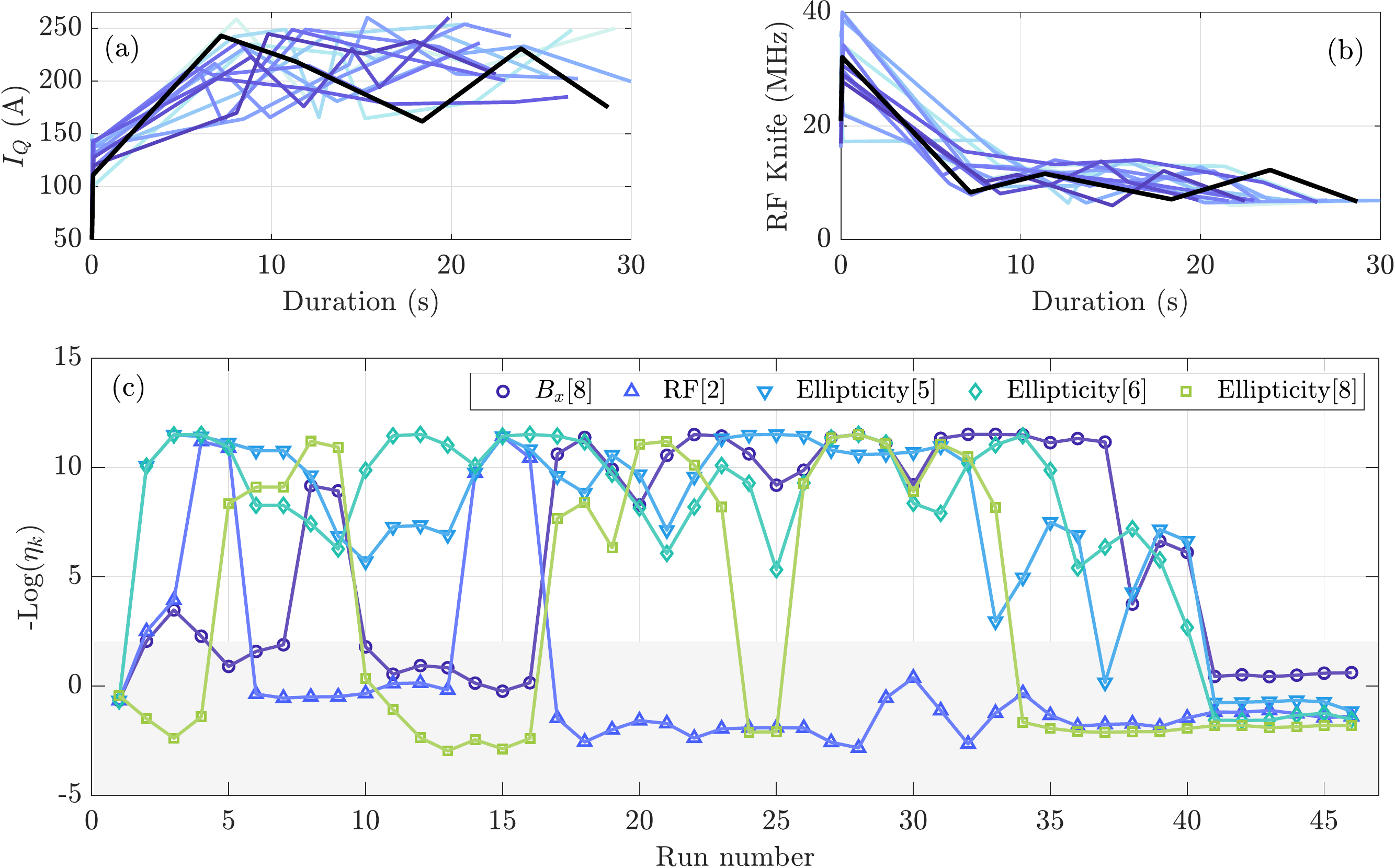}
	\caption{\label{fig:Length_Scales}(a) \& (b) illustrate the progression of the quadrupole current $I_Q$ and RF knife frequency settings during the TOP substages, respectively, as produced during the optimization. The settings are plotted against the duration of the evaporative cooling stage. Darker colours indicate settings produced later into the optimization procedure. The minimum cost is achieved with the settings shown in black. (c) shows the evolution of the 5 most sensitive $\eta_k$ as more data is added during the optimization routine. The settings shown are indexed by a number which indicates the substage of the evaporative cooling stage, as given in table~\ref{fig:EvapArray}. The shaded area indicates the region of $-\mathrm{Log}(\eta_k)<2$ within which a setting is deemed sensitive.}
\end{figure}

We utilize the cost landscape fitted by the GP learner to determine the sensitivity of the different settings. As detailed in section~\ref{Sec:GP}, the variables $\eta_k$ provide a measure of how steeply the cost function varies about the predicted minimum with respect to each setting. As a heuristic indicator of sensitivity, we define a setting to be sensitive if the associated $\eta_k$ is greater than $\mathrm{exp}(-2)$. Figure~\ref{fig:Length_Scales} illustrates the convergence of $\eta_k$ as more data is added during the evaporative cooling optimization. For clarity, only the five most sensitive settings are shown.

We find that the cost is highly sensitive to the final amplitude of the magnetic field in the TOP trap ($B_x$[7]), as well as the initial (RF[1]) and final (RF[7]) radiofrequencies of the knife. This can be understood as follows: the initial frequency determines the threshold energy above which atoms are ejected from the trap; this frequency must be sufficiently high so as not to immediately cut away a large number of atoms when the RF knife is first turned on when evaporation begins. A combination of final knife frequency and TOP amplitude determines the final, lowest energy cut in the evaporation ramp. If this is too high, the cloud is hotter and fewer atoms accumulate within the region-of-interest after time-of-flight. If this is too low, the evaporation sequence unneccesarily ejects atoms which would otherwise have contributed to the BEC component. The cost is also sensitive to the final RF knife cut in the quadrupole trap, as this determines the temperature of the atomic cloud when it is loaded into the TOP trap.

Surprisingly, we find the cost is highly sensitive to the TOP field ellipticity during certain substages. This setting was fixed to 0 during previous manual optimization, as this was expected to yield the best results. From observations of cloud positions when trapped in the quadrupole or TOP trap, we have determined that the rotation axis of the TOP field is not perfectly aligned with the symmetry axis of the quadrupole field, which increases the displacement between the energy minima of an atom in these two traps. In addition to any centre-of-mass motion of the cloud, which may be induced as the cloud is transferred from the quadrupole to the TOP trap, other multipole oscillations in the cloud may be excited. We postulate that a non-zero ellipticity in the TOP field provides an asymmetric confinement force, which may help to eradicate or damp excitations in the cloud that would otherwise affect the efficiency of subsequent evaporative cooling. The ellipticity of the TOP field can also help to counter-balance any asymmetry in the quadrupole field. These unexpected results produced by the optimization procedure, which at first seem counterintuitive, can provide hints to the experimentalist as to where imperfections might exist in the apparatus.

\subsection{Multiple stage optimization of laser cooling and evaporative cooling}

\begin{figure}[ht]
	\centering
	\includegraphics[width = \linewidth]{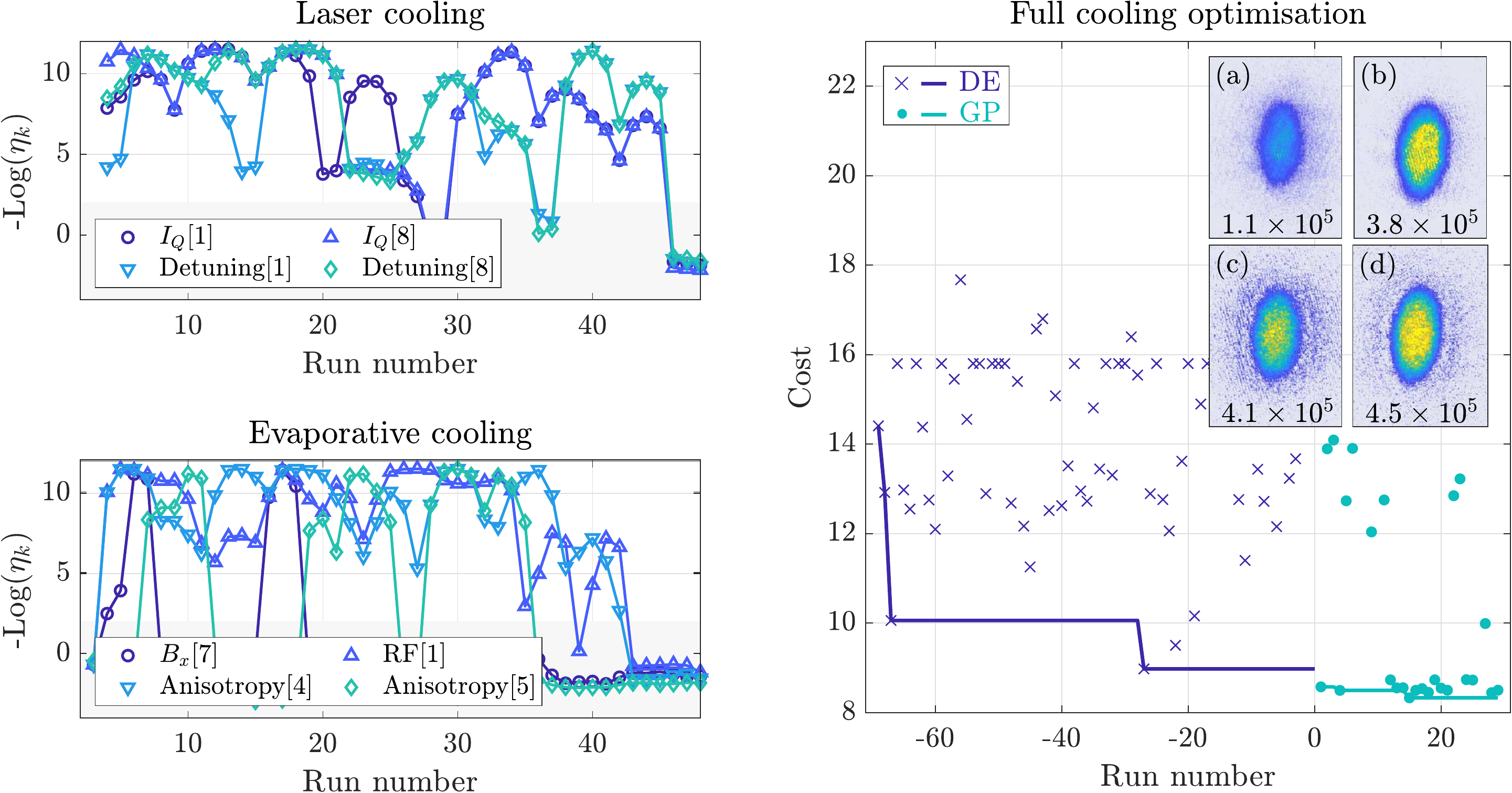}
\caption{Cost vs. run number for optimization of all cooling processes, using the GP method. Laser cooling panel: evolution of the 4 most sensitive $\eta_k$ during optimization of laser cooling (cMOT). Evaporative cooling panel: evolution of the 4 most sensitive $\eta_k$ during optimization of evaporative cooling. Right: The cost versus run number for the overall cooling optimisation, using the 18 most sensitive settings, for the initial DE method (\textcolor[rgb]{0.2422,0.1504,0.6603}{\textbf{$\times$}}) and the subsequent GP method (\textcolor[rgb]{0.0433,0.7411,0.7394}{$\bullet$}). The moving minimum cost is illustrated by the solid line. Sequences used in building the training set using the DE method are again labelled with negative numbers. Insets: (a) - manually-optimized BEC, (b) - BEC following evaporative cooling optimization, (c) - BEC following laser cooling optimization and (d) - BEC following the full cooling optimization. The temperature, as extracted from a fitting procedure of (d) is 69(6) \si{\nano\kelvin}. Images use the same colour scheme as that used in figure~\ref{fig:Absorption} and values within the insets correspond to the total atom numbers.}
\label{fig:MultiStageLength}
\end{figure}

The previous section shows that GP regression is the most rapidly converging of the methods tested in our experimental context. For the remainder of this paper, we therefore focus on this method. We use the GP method to optimize the cMOT stage, approximately doubling the number of laser cooled atoms produced. In figure ~\ref{fig:MultiStageLength}, the Laser cooling panel illustrates the convergence of the four most sensitive $\eta_k$ of this stage as the optimization progresses.

The number of sequences required for the GP method to converge increases with the number of settings. In addition, the computation time scales as the cube of the number of costs over which the GP fits. Given this, it is advantageous to reduce the number of settings as far as is reasonable without compromising the outcome of the optimization. Determining the most sensitive settings allows simultaneous optimization of all cooling stages (laser cooling in the cMOT, evaporative cooling in the quadrupole trap and TOP traps) in a reasonable time. We again utilize the DE method to produce a training set of $2M = 36$ settings/cost pairs. The values of insensitive settings in the laser cooling and evaporative cooling stages were fixed to the best values found during the separate optimization of each stage. 

Using the GP learner and by optimizing only the sensitive settings, we are able to produce a BEC from random initial settings after only 12 experimental sequences (following the 36 runs used to build the training set). The optimization produces a BEC with an atom number of $4.5 \times 10^5$, which is greater than atom numbers produced in the optimization of the cooling stages separately. Figure~\ref{fig:MultiStageLength}, (a-d), illustrates the improvements in BEC atom number after we have optimized the stages individually and collectively. This faster optimization routine, using only the sensitive settings, can be used to perform quick and regular re-optimization to keep an experimental apparatus tuned up to the best of its capability.

\subsection{Tailoring the cost function}

A maximized atom number in the BEC is often desirable and this motivated our earlier choice of cost function. 
However, depending on the scenario, other quantities may be of greater importance. 
For example, when performing alignment of optical elements, it is more useful to maximize the repetition rate of the experiment. 

\subsubsection{Minimizing sequence duration.}

We use our optimization routine to find settings which produce a BEC of a threshold atom number in the shortest possible time.
In general, BEC experiments have sequence durations ranging from a few seconds to minutes, depending on the implementation details of each apparatus. 
We use the cost function $$f(\tilde{N}) = -\left( 1 + \mathrm{arctan}(\tilde{N}-\tilde{N}_{0}) \right) / \left( 1 + \tilde{t} \hspace{1pt} \right) \hspace{5pt}$$ where $\tilde{N}_{0}$ is a threshold number of atoms within our region-of-interest, which we choose to correspond to an overall BEC size of $1 \times 10^5$ atoms, and $\tilde{t}$ is the sequence duration. 
This cost function rewards a short sequence time and penalizes settings which do not produce a BEC of a threshold atom number; there is also little reward for producing a BEC with an excess of atoms. 
With no other changes to the optimization routine, the optimized settings produce a BEC of $9.6 \times 10^4$ atoms and reduce our overall sequence time from \SI{58}{\second} to \SI{46}{\second}, a time saving of over \SI{20}{\percent}. This demonstrates the power of online optimization to reconfigure an apparatus to achieve the aims of the user.

\subsubsection{Minimising temperature.}
\label{sec:minT}
\begin{figure}[ht]
	\centering
	\includegraphics[width = \linewidth]{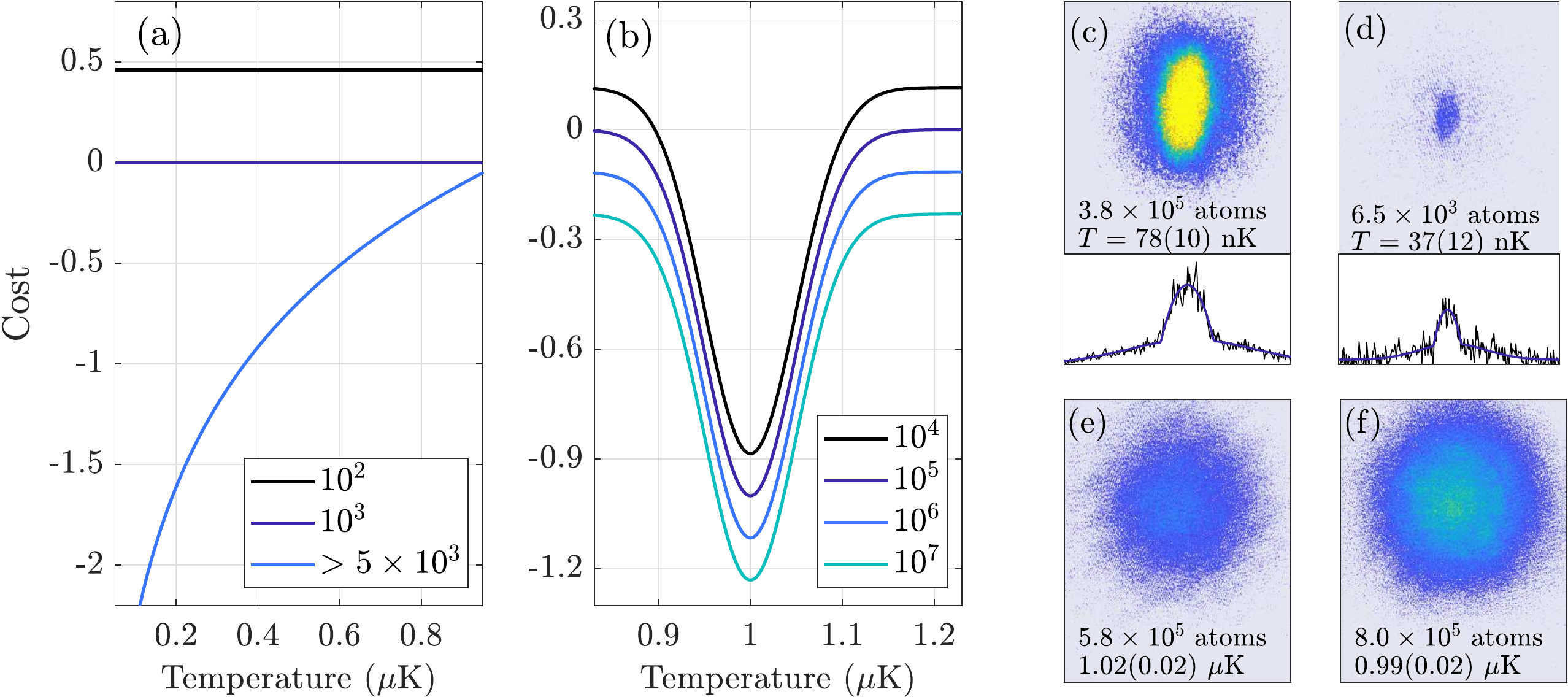}
	\caption{Results of optimizations detailed in sections~\ref{sec:minT} \& \ref{sec:maxNspecT}. (a-b) Cost vs. atom number and temperature for the cost functions presented in sections~\ref{sec:minT} \& \ref{sec:maxNspecT}, respectively. Values in the legend of each plot indicate the total atom number of the cloud. (c-d, upper) Absorption images of BECs produced using the optimization routines presented in section~\ref{sec:evapCooling} \& \ref{sec:minT}, respectively. (c-d, lower) Bimodal fits (purple) to the atomic density profiles (black) extracted from the absorption images above. (e-f) \SI{1}{\micro\kelvin} clouds produced using a truncated version of the evaporative cooling routine presented in section~\ref{sec:evapCooling} and by the optimization presented in \ref{sec:minT}, respectively.}
	\label{fig:CombinedTemp}
\end{figure}

We use our machine learner to find settings which minimize the temperature of the ultracold gas.
Temperature cannot be made arbitrarily low, as $N \to 0$ for $T \to 0$ for an evaporative cooling process, so we incorporate a threshold number $N_0$ into the cost function.
We define the cost function $f(N,T)$ to be

\begin{equation}
\begin{aligned}
\mathrm{if} \hspace{10pt} & N > N_{\mathrm{0}} \\
&\hspace{10pt} f(N,T) = \mathrm{log}(T) \\
\mathrm{else} & \\
&\hspace{10pt} f(N,T) = - 0.2 \hspace{3pt} \mathrm{log}(N)
\end{aligned}
\end{equation} 

\noindent where $T$ is the cloud temperature and $N$ is the total atom number in the atomic cloud. 
Figure~\ref{fig:CombinedTemp} (a) illustrates this cost as a function of atom number and temperature. 
For sufficient atom numbers, the cost depends only on temperature and encourages the learner to reduce $T$.
For smaller atom numbers, the fits from which temperature is inferred can fail, and so only the atom number itself is used to determine the cost. 
We take $N_{\mathrm{0}} = 5 \times 10^3$ as a lower bound for the number of atoms in the BEC. 
The prefactor of 0.2 minimizes the discontinuity in the cost either side of the threshold atom number, which assists our gradient-based method in finding the optimal parameter set.

As shown in figure~\ref{fig:CombinedTemp} (d), optimizing for temperature produces a \SI{37(12)}{\nano\kelvin} cloud of around $6.5 \times 10^3$ atoms. 
The resulting cloud is significantly colder than that produced using the optimization detailed in section~\ref{sec:ML_methods} (figure~\ref{fig:CombinedTemp} (c)).  

\subsubsection{Maximising atom number at a specific temperature.}
\label{sec:maxNspecT}

Other optimizations could also be conceived, such as maximizing the atom number of a thermal cloud at a specified temperature. We use the cost function $$f(N,T) = - \mathrm{exp}(-(T/T_0 - 1)^2) - \mathrm{log}(N) \hspace{5pt} ,$$ where $T$ is the temperature, $T_0 = \SI{1}{\micro\kelvin}$ is the target temperature and $N$ is the total atom number. This cost function favours the production of a \SI{1}{\micro\kelvin} cloud with the greatest atom number, and is illustrated in figure~\ref{fig:CombinedTemp} (b) as a function of temperature for different atom numbers. 

Figure~\ref{fig:CombinedTemp} (f) shows an image of a thermal cloud produced by settings optimised in this way and, for comparison, figure~\ref{fig:CombinedTemp} (e) shows a \SI{1}{\micro\kelvin} cloud produced by truncating the evaporative cooling ramp from table~\ref{fig:EvapArray}.
The \SI{1}{\micro\kelvin} cloud produced through this optimization has an atom number that is \SI{38}{\percent} larger.
Another variant of this scheme could be to minimize $T$ for a thermal gas with a specified atom number, which would be possible with only a minor adjustment to the cost function described above.

\section{Conclusion}

The value of machine learning in finding patterns and optima in data which depends on many parameters is apparent across multiple fields of research~\cite{Kalantre2019}. In our specific case, machine learning has provided a means for autonomous experimental optimization. We have compared the convergence rate of three optimization methods. Most notably, for the first time, we have optimized all cooling stages involved in a quantum gas experiment simultaneously. The optimization is quick and achieves our aim of increasing the atom number in a BEC, which is beneficial for improved signal-to-noise ratios when measuring atom numbers in future experiments.

We have used the GP method to identify the sensitive settings within each cooling stage. By restricting the attention of the learner to only consider these sensitive settings, it becomes possible to optimize the experiment as a whole with only a small number of sequences. Optimisation can be performed within an hour, allowing daily optimisation if necessary to maintain peak performance for producing consistent, high-quality data. Long-term drifts which would otherwise degrade the apparatus' performance can thus be easily mitigated, by scheduling regular optimization routines, e.g. once a week.

Certain features of our optimal solutions are counterintuitive: improvements arising from an elliptical TOP field during the evaporative cooling stage were not expected and would not generally be explored by a researcher. These features may indicate underlying physics, or may allude to the presence of imperfections in the experimental apparatus.

One caveat is that the point of convergence, or optimum, may be one for which the length scale of any parameter is extremely short. While we hope to find the global minimum of the cost function, it is of little experimental value if a perturbation from the prescribed experimental settings leads to a sharp response in the cost. The stability of the solution can be evaluated by assessing the average cost over multiple runs for each input and building separate models for both $\E{[Y]}$ and $\Var{[Y]}$. These can be jointly optimized to produce a solution which not only works to achieve the user's optimization aim but also reduces shot-to-shot fluctuations which limit the resolution of an experiment. In the interest of short optimization routines, we have decided against this approach. We have observed that the optima found are no less stable than the previous, manually optimized values. Even so, shorter optimization rountines can be performed more frequently to counter long-term drifts.

Given the desirability of short optimization routines, and as illustrated by the relative rates of convergence between the methods, we conclude that the GP regression method is of greatest utility in our experimental context. Our optimization routine produces a relatively small amount of training data which, consequently, may reduce the suitability of an ANN-based method, as these typically require many thousands of data points to accurately train the network weights.

\section*{Acknowledgements}
The authors would like to thank Henry Howard-Jenkins for useful discussions. This work was supported by the EPSRC Grant Reference EP/S013105/1. We gratefully acknowledge the support of NVIDIA Corporation with the donation of the Titan Xp GPU used for parts of this research. AJB, KL and DG thank the EPSRC for doctoral training funding. The data that support the findings of this study are available from the corresponding author upon reasonable request.

\appendix
\section{Further detail on the optimization cost function}
\label{sec:app}

In our context, the intention of the evaporative cooling stage is to increase the phase-space density (PSD) \cite{PethickSmith} of an atomic cloud to the critical value required for Bose-Einstein condensation. As detailed in sec.~\ref{Sec:evap}, evaporative cooling is performed by ejecting atoms of higher-than-average energy. The remaining atoms rethermalize to form a colder atomic cloud with a momentum distribution that is more strongly peaked around $k=0$, where $k$ is atomic momentum, with the onset of a macroscopic occupation of $k\approx 0$ at the critical point for Bose-Einstein condensation. Although the temperature of the atomic cloud is favourably reduced, the mechanism of evaporative cooling reduces the total number of atoms as the stage progresses.

Our optimization objective, as described in sec.~\ref{sec:ML_methods}, is to maximize the number of atoms within a small circular region centred on the atomic cloud after time-of-flight (TOF) expansion. TOF expansion is a popular method of extracting the momentum distribution of an atomic cloud~\cite{PethickSmith}. The bimodal density distribution for a BEC after TOF expansion is well known~\cite{Davis1995}: the thermal cloud expands to form a broad Gaussian pedestal, while the BEC forms a parabolic profile in the centre. This is illustrated in fig.~\ref{fig:Absorption}. Our counting region captures the number of atoms with momentum $k$ less than a threshold $k_c$, where $k_c$ is set by the radius of the region. To ensure our region includes mostly the BEC component, the radius of the counting region is chosen to be comparable to the Thomas-Fermi radius of a BEC of 〖$10^5$ atoms in a trap with similar parameters to our manual-optimized BEC sequence. Fig.~\ref{fig:Absorption} illustrates that this region predominantly includes the BEC component of the cloud for both the manually-optimized and machine learning-optimized BECs.

\begin{figure}
	\centering
	\includegraphics[width = \linewidth]{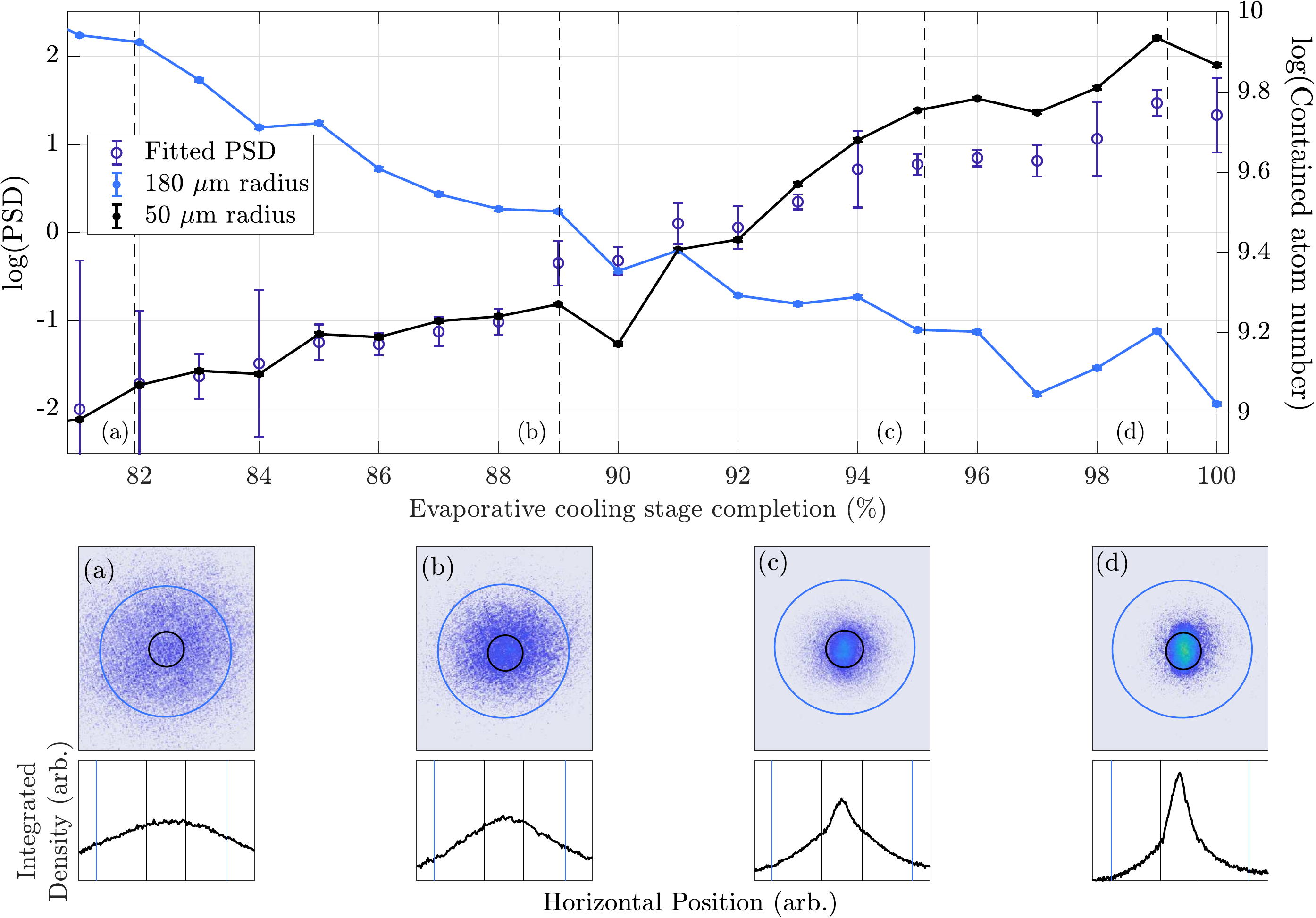}
	\caption{Fitted PSD (purple) as a function of evaporative cooling stage progression. Similarly, the number of atoms contained within regions of size \SI{180}{\micro\metre} (blue) and \SI{50}{\micro\metre} (black) as a function of evaporative cooling stage progression for the same data. Images (a-d, upper) show atomic density distributions after TOF for 82\%, 80\%, 95\% and 99\% stage completion, respectively. The counting regions of radius \SI{180}{\micro\metre} (blue) and \SI{50}{\micro\metre} (black) are also illustrated. (a-d, lower) show the density distribution integrated in the vertical axis plotted against horizontal position. The horizontal extent of the \SI{180}{\micro\metre} (blue) and \SI{50}{\micro\metre} counting regions are illustrated by blue and black lines, respectively.}
	\label{fig:PSDcomp}
\end{figure}

We vary the completion percentage of the evaporative cooling stage and extract the PSD from a bimodal fit. As shown in fig.~\ref{fig:PSDcomp}, the fitted PSD (purple) rises to and above the critical value for Bose-Einstein condensation. We also show the number of atoms contained within a region much larger than the extent of the BEC (\SI{180}{\micro\metre} radius, blue) and a region of comparable size (\SI{50}{\micro\metre} radius, black). We divide the number of atoms contained within the larger region by 8 for easier comparison with the smaller region on the same axis.

The number of atoms contained within the smaller region increases with PSD, as the atomic ensemble re-establishes a momentum distribution more strongly peaked around $k=0$. Additionally, and as expected, the number of atoms contained within the larger region decreases as the stage progresses and as the total atom number is reduced. As shown for this data set, extracting PSD from a fitting procedure would provide the equivalent information to a learner as our atom counting method. Nevertheless, when the atomic cloud is faint or not visible, as is the case for earlier stages of the evaporative cooling stage, the fit fails. In contrast, the counting region method relies on fewer parameters and is significantly more robust to small or faint clouds, which are encountered regularly during an optimization routine.

\section*{References}
\bibliographystyle{iopart-num}
\bibliography{library}

\end{document}